\documentclass[english,10pt,a4paper,final]{article}
\usepackage[english]{babel}
\usepackage{subfig}
\usepackage{float}
\usepackage[utf8]{inputenc}
\usepackage{graphicx}
\usepackage{color}
\usepackage{xfrac}
\usepackage{titling}
\usepackage{mathtools}
\usepackage{lmodern}

\usepackage[super,numbers,sort&compress]{natbib}
\usepackage{float}

\RequirePackage{lastpage}  

\RequirePackage[left=1.7cm,%
                right=1.7cm,%
                top=2.2cm,%
                bottom=2.5cm,%
                headheight=12pt,%
                letterpaper]{geometry}%
                
\RequirePackage[labelfont={bf,sf},%
                labelsep=period]{caption}

\RequirePackage[colorlinks=true, allcolors=blue]{hyperref}

\usepackage{amsmath}
\usepackage{appendix}
\usepackage{amssymb}
\usepackage{enumerate}
\usepackage{multicol}
\usepackage{gensymb}
\usepackage{braket}
\usepackage{bm}

\usepackage{caption,mathtools,longtable}       
\usepackage{array}  

\newcommand{\rb}{\textbf{r}}

\renewcommand{\u}{\mathcal{U}}

\title{\LARGE\bf Atomtronic protocol designs for NOON states}
\author{Daniel S. Grün$^1$, Karin Wittmann W.$^1$, Leandro H. Ymai$^2$, $^*$Jon Links$^3$ \& Angela Foerster$^1$
\vspace{0.2cm} \\
\small{$^1$ Instituto de Física da UFRGS, Av. Bento Gonçalves 9500, Agronomia, Porto Alegre, 91501-970, RS, Brazil}\\
\small{$^2$ Universidade Federal do Pampa, Av. Maria Anunciação Gomes de Godoy 1650, Malafaia, Bagé, 96413-170, RS, Brazil}\\
\small{$^3$ School of Mathematics and Physics, The University of Queensland, Brisbane, QLD, 4072, Australia} \\
\small{$^*$Corresponding author} }
\date{\vspace{-5ex}}

\begin{document}
\maketitle
\begin{abstract}
The ability to reliably prepare non-classical states will play a major role in the realization of quantum technology.
NOON states, belonging to the class of Schr\"odinger cat states, have emerged as a leading candidate for several applications.
Starting from a model of dipolar bosons confined to a closed circuit of four sites, we show how to generate NOON states. 
This is achieved by designing protocols to transform initial Fock states to NOON states through use of time evolution, application of an external field, and local projective measurements. 
By variation of the external field strength, we demonstrate how the system can be controlled to encode a phase into a NOON state. 
We also discuss the physical feasibility, via an optical lattice setup. Our proposal illuminates the benefits of quantum integrable systems in the design of atomtronic protocols.
\end{abstract}

\begin{multicols}{2}

\noindent
Quantum systems are widely considered to be the most promising foundation
for the next generation of platforms in computing, communication, measurement and simulation. 
This is primarily due to the
properties of state superposition and entanglement. To realize the potential for 
progress, it is necessary to establish protocols that are capable of generating important quantum states.

The NOON state is a fundamental example. It is
an ``all and nothing'' superposition of two different modes  \cite{rosetta, afek}.  For $N$ particles, it has the form
\begin{align}
|{\rm NOON}\rangle  = \frac{1}{\sqrt{2}} \left(|N,0\rangle 
+ e^{i\varphi} |0,N\rangle\right)
\label{noon} 
\end{align}
where the phase $\varphi$ typically records information in applications. These include: in the fields of quantum metrology and sensing, performing precision phase-interferometry at the Heisenberg limit\cite{bollinger1996} and overcoming diffraction limits in quantum lithography\cite{boto2000}; in tests of fundamental physics, NOON states are used to study Bell-type inequalities violation\cite{wildfeuer2007}; they offer promising  applications in Quantum Communication and Quantum Computing\cite{pan2012}, and their utilization is expected to extend to areas such as chemistry and biology\cite{haas2018}. After an early success, using photon pairs and Hong-Ou-Mandel (HOM) interferometry\cite{rarity1990}, several schemes have followed for the production and detection of photonic NOON states \cite{afek, white1, white2,kamide2017, esguibar2019}. There are also proposals using other architectures, such as circuit QED\cite{merkel2010}, trapped ions \cite{hu2012}, and  Bose-Einstein condenstates\cite{cable2011}. 

The atomtronic creation of Bose-atom NOON states would enable new tests, using massive states, of the foundations for quantum mechanics. One step in this direction is a proposal to demonstrate the matter-wave equivalent of the HOM effect \cite{lewis_swan}. 
Prospects for creating Bose-atom NOON states using  a double-well potential were first floated some time ago \cite{cirac}. This early work considered an attractive system, which is prone to instability. In principle a more robust repulsive system can be prepared to evolve to a high-fidelity approximation of a NOON state. 
However, the drawback there is that the process is associated with an extremely large time scale. Recently, new studies of the double-well system have been undertaken to reduce the time scale. One example proposes to adiabatically vary the system parameters through an excited-state phase transition during the process \cite{bychek18}. Another study employs periodic driving to lower the NOON-state evolution time \cite{vanhaele2020}.
Nonetheless, the time to generate a NOON state in these examples still, increasingly, scales with the total number of particles.

\begin{figure}[t!]
    \centering
             \includegraphics[width=13cm]{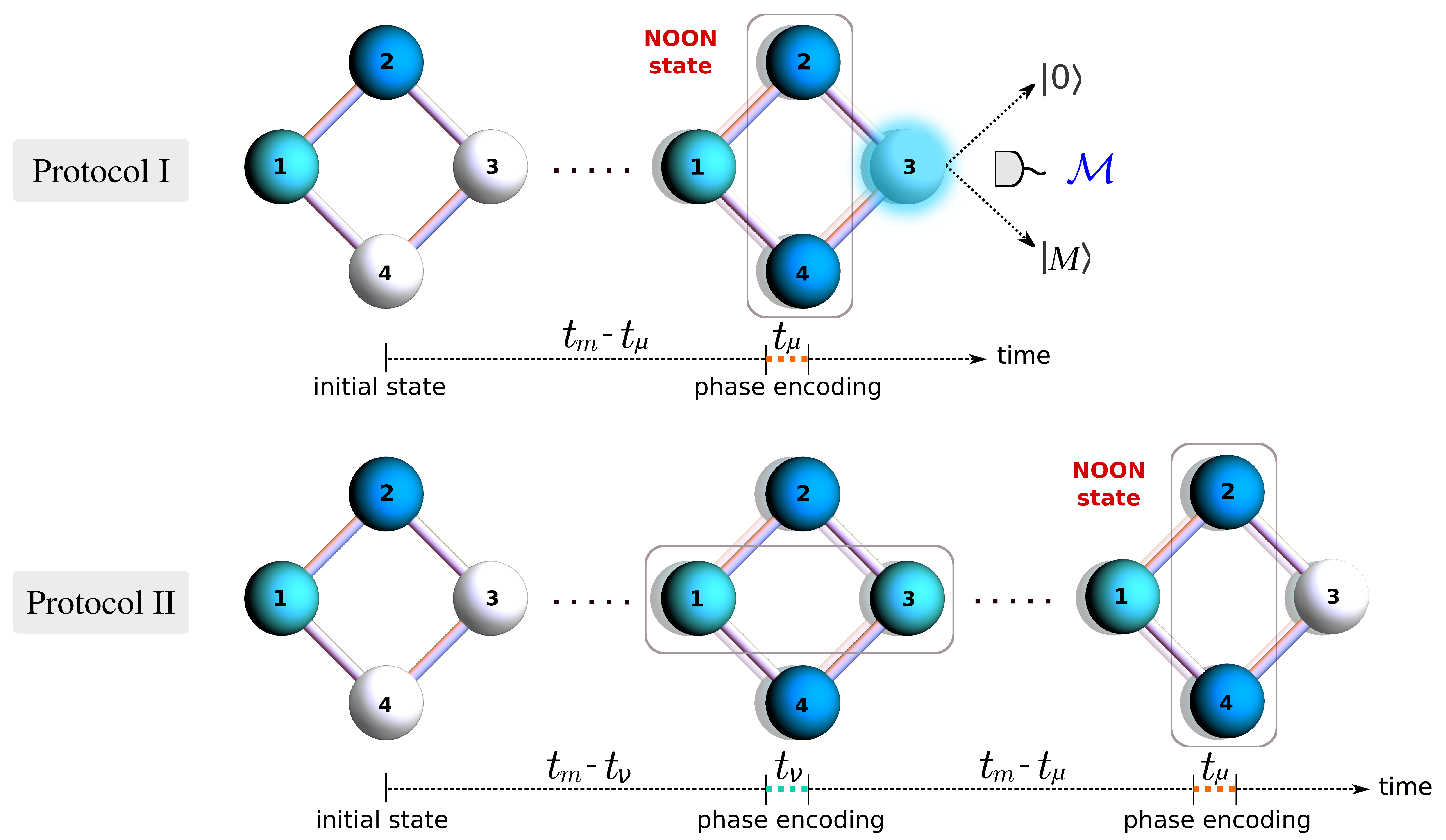}
    \caption{NOON state generation scheme. The four circles on the left represent the initial state, with white indicating an empty site,  cyan and blue  corresponding to $M$ and $P$ particles respectively. The solid lines connecting the circles denote tunneling between nearest neighbor sites. Rectangles represent applied external fields to sites 1-3 and 2-4. In Protocol I, the system initially evolves during a time of $t_{\text{m}} - t_\mu$. Then, an applied field across sites 2-4 is switched on, and a phase is encoded during the time $t_\mu$. Finally, the light blue halo portrays a projective measurement process at site 3, denoted by $\mathcal{M}$, resulting in two possible NOON states across sites 2-4. In Protocol II, the system first evolves during a time of $t_{\text{m}} - t_\nu$. Then, an applied field to sites 1-3 is switched on for time $t_\nu$. Next, the system evolves for $t_{\text{m}} - t_\mu$, after which the applied field to sites $2-4$ is switched on. This results in a NOON state across sites 2-4 without performing a measurement procedure.} 
    \label{fig:both-protocols}
\end{figure}

Here we present an alternative to circumvent these issues. Our approach adopts a  closed-circuit of four sites, with a Fock-state input of $M$ particles in site 1, $P$ particles in site 2, and no particles in sites 3 and 4, denoted as $|\Psi_0\rangle=|M,P,0,0\rangle$.  The initial step is to create an uber-NOON state, with the general form
\begin{align*}
|\text{u-NOON}\rangle  &= \frac{1}{2} \left(|M,P,0,0\rangle + e^{i\varphi_1} |M,0,0,P\rangle \right.  \\
&\qquad \left. 
+e^{i\varphi_2}|0,P,M,0\rangle  
+e^{i\varphi_3} |0,0,M,P\rangle
\right)
\end{align*}
for a set of phases $\{\varphi_1,\,\varphi_2,\,\varphi_3\}$. This state may be viewed as an embedding of NOON states (\ref{noon}) within two-site subsystems. We then describe two protocols to extract a NOON state from an uber-NOON state, one through dynamical evolution followed by local projective measurement and post-selection, the other from dynamical evolution alone. The protocols are schematically presented in Fig. \ref{fig:both-protocols}. 

The approach taken has the following properties:  
(i)
The system has long-ranged interactions, described
by the Extended Bose-Hubbard Model (EBHM)\cite{goral2002}. There exists a choice of the coupling parameters for which this model  is integrable\cite{tyfl}. As in other physically realized integrable systems \cite{kinoshita2006, liao2010, pagano2014, batchelor2016, yang2017, breunig2017, wang2018, lmg2020}, this property facilitates several analytic calculations for physical quantities. Here, integrability exposes the protocols available for NOON state generation. The execution time is found to be dependent on the difference between the two initially populated sites within the four-site system. It is independent of total particle number, offering an encouraging prospect for scalability.  
\noindent
(ii) The system can be controlled by breaking the integrability over small time scales. Encoding of the phase into a NOON state only requires breaking of integrability over an interval that is several orders of magnitude smaller than the entire execution time. This causes minimal loss in fidelity.   
(iii) With currently available  technology, the system may be realized and controlled using dipolar atoms (e.g. dysprosium or erbium) trapped in an optical lattice\cite{dy, er}. In this setup, the evolution times that we compute for NOON-state generation are of the order of seconds.

For the four-site configuration, the EBHM Hamiltonian is
\begin{flalign}
\begin{split}
    H = &\frac{U_0}{2}\sum_{i=1}^{4} N_i(N_i-1) + \sum_{i=1}^{4}\sum_{j=1, j\neq i}^{4}\frac{U_{ij}}{2}N_i N_j\\
    &- \frac{J}{2}\big[(a_1^\dagger + a_3^\dagger)(a_2 + a_4) + (a_1 + a_3)(a_2^\dagger + a_4^\dagger)\big],\\
\end{split}
\label{eq:extended-bh}
\end{flalign}
where $a_j^\dagger,\,a_j$ are the creation and annihilation operators for site $j$, and $N_j=a_j^\dagger a_j$ are the number operators.  The total number operator $N=N_1+N_2+N_3+N_4$ is conserved. Above, $U_0$ characterizes the interaction between bosons at the same site, $U_{ij} = U_{ji}$ is related to the long-range (e.g. dipole-dipole) interaction between bosons at sites $i$ and $j$, and $J$ accounts for the  tunneling strength between different sites. 

Below, we describe two protocols that enable the generation of NOON states,
with fidelities greater than  0.9.
A physical setup to implement them, drawn on currently available technology, is discussed. 
\end{multicols}

\begin{multicols}{2}

\section*{Results}
Insights into the physical behaviour of Eq. (\ref{eq:extended-bh}) become accessible at integrable coupling. Setting $U_{13} = U_{24} = U_0$, the system acquires two additional conserved quantities, $Q_1$ and $Q_2$, such that $2Q_1 = N_1 + N_3 - a_1^\dagger a_3 - a_1 a_3^\dagger$ and $2Q_2 = N_2 + N_4 - a_2^\dagger a_4 - a_2 a_4^\dagger$. Together with the total number of particles $N$ and the Hamiltonian $H$, the system possesses four independent, conserved quantities. This is equal to the number of degrees of freedom, satisfying the criterion for integrability. Suppose that, initially, there are $M$ atoms in site $1$ and $P$ atoms in site $2$. We identify the resonant tunneling regime as being achieved when $U|M-P| \gg J$ (see Methods for details), where $U = (U_{12}-U_{0})/4$. This regime is characterized by sets of bands in the energy spectrum (see Supplementary Note 1). In this region, an effective Hamiltonian $H_{\rm eff}$ enables the derivation of analytic expressions for several physical quantities.

In the settings discussed above, the system described by Eq. (\ref{eq:extended-bh}) provides the framework to generate uber-NOON states when $N=M+P$ is odd \cite{preprint}. To encode phases, however, it is necessary to break the integrability in a controllable fashion. Here, we introduce two idealized protocols to produce NOON states with general phases by breaking the system's integrability with externally applied fields. We call the subsystem containing sites $1,3$ as $A$, and the one containing sites $2,4$ as $B$. We denote three time intervals: $t_{\text{m}}$, $t_\mu$ and $t_\nu$. The first, corresponding to integrable time evolution, is associated with evolution to a particular  uber-NOON state. 
The others, associated with smaller scale  non-integrable evolution, produce phase encoding. Both protocols are built around a general time-evolution operator 

\begin{equation*}
    \u(t,\mu,\nu) = \exp\left(-\frac{it}{\hbar}[H + \mu(N_2-N_4) 
    + \nu(N_1-N_3)]\right),
\end{equation*}
where the applied field strengths $\mu$, $\nu$ implement the breaking of integrability. 
It is convenient to introduce the phase variable $\theta=2\mu t_\mu/\hbar$, and to fix $ t_\nu=\hbar \pi/(4M\nu)$, with $\hbar$ the reduced Planck constant. 

\subsection*{Protocol I}
In this protocol we employ breaking of integrability through an applied field to subsystem $B$ and a measurement process. The protocol consists of three sequential steps, schematically depicted in Fig. \ref{fig:both-protocols}:

\begin{enumerate}
    \item[(i)] $\ket{\Psi_{1}^{\text{I}}} = \u(t_{\text{m}}- t_\mu, 0, 0)\ket{\Psi_0}$;
    \item[(ii)] $\ket{\Psi_{2}^{\text{I}}} = \u(t_\mu, \mu, 0)\ket{\Psi_{1}^{\text{I}}}$;
    \item[(iii)] $\ket{\Psi_{3}^{\text{I}}} = \mathcal{M}\ket{\Psi_{2}^{\text{I}}}$,
\end{enumerate}
where $t_\text{m}=\hbar \pi/(2\Omega)$ (see Methods)  and $\mathcal{M}$ represents a projective measurement of the number of bosons at site 3 (which could be implemented, in principle, through Faraday rotation detection\cite{yamamoto, schafer}). A measurement outcome of 0 or $M$ heralds a high-fidelity NOON state in subsystem $B$. For other measurement outcomes, the output is discarded and the process repeated (post-selection). 

\subsubsection*{Idealized limit}
There is an idealized limit for which the above protocol has perfect success probability and output fidelity.
Taking $t_\mu\rightarrow 0$, $\mu\rightarrow\infty$  such that $\theta$ remains finite, and using the effective Hamiltonian, provides explicit expressions for the uber-NOON states that result at steps (i) and (ii)
\begin{flalign}
\begin{split}
    & \hspace{-0.4cm}\ket{\Psi_{1}^{\text{I}}} = \frac{1}{2}\Big(\beta\ket{M,P,0,0} + \ket{M,0,0,P}\\
    &\phantom{\hspace{-0.4cm}\ket{\Psi_{1}^{\text{I}}} = \frac{1}{2}\Big(} + \ket{0,P,M,0} - \beta\ket{0,0,M,P}\Big)\\
    &\hspace{-0.4cm}\ket{\Psi_{2}^{\text{I}}} = \frac{1}{2}\Big(\beta\ket{M,P,0,0} + e^{iP\theta}\ket{M,0,0,P}\\
    &\phantom{\hspace{-0.4cm}\ket{\Psi_{2}^{\text{I}}} = \frac{1}{2}\Big(} + \ket{0,P,M,0} - \beta e^{iP\theta}\ket{0,0,M,P}\Big)\\
\end{split}
\label{eq:psi12-jon}
\end{flalign}
Note that due to the conservation of $N_1+N_3$ and $N_2+N_4$ under the effective Hamiltonian, Fock states such as $\ket{M,0,P,0}$ and  $\ket{0,M,0,P}$ do not appear in the above expression. Next, the two possible states at step (iii) depend on the measurement outcome $r$ at site 3:
\begin{equation}
    \ket{\Psi_{3}^{\text{I}}} = 
    \begin{cases}
     \displaystyle 
    \frac{1}{\sqrt{2}}\left(\beta\ket{M,P,0,0} + e^{iP\theta}\ket{M,0,0,P}\right), \;r = 0,\\
     \displaystyle
    \frac{1}{\sqrt{2}}\left(\ket{0,P,M,0} - \beta e^{iP\theta}\ket{0,0,M,P}\right), \;r = M,
    \end{cases}
    \label{eq:psi3jon}
\end{equation}
with $\beta = (-1)^{(N+1)/2}$. These states are recognized as products of a NOON state for subsystem $B$ with Fock basis states for subsystem $A$. 

In the non-ideal case with non-zero $t_\mu$ and finite $\mu$, there is a small probability that the measurement outcome $r$ is neither 0 or $M$. Numerical benchmarks for the measurement probabilities and NOON state output fidelities are provided in a later section. Next, we describe a second protocol.    

\subsection*{Protocol II}
Now we specify an alternative protocol that does not involve measurements, so post-selection is not required. 
Employing the same initial state $\ket{\Psi_0}$, the following sequence of steps are implemented to arrive at a NOON state in subsystem $B$ (illustrated in Fig. \ref{fig:both-protocols}):
\begin{itemize}
    \item[(i)] $\ket{\Psi_{1}^{\text{II}}} = \u(t_{\text{m}} - t_\nu, 0, 0)\ket{\Psi_0}$;
    \item[(ii)] $\ket{\Psi_{2}^{\text{II}}} = \u(t_\nu, 0, \nu)\ket{\Psi_{1}^{\text{II}}}$;
    \item[(iii)] $\ket{\Psi_{3}^{\text{II}}} = \u(t_{\text{m}} - t_\mu, 0, 0)\ket{\Psi_{2}^{\text{II}}}$;
    \item[(iv)] $\ket{\Psi_{4}^{\text{II}}} = \u(t_\mu, \mu, 0)\ket{\Psi_{3}^{\text{II}}}$.
\end{itemize}

\subsubsection*{Idealized limit}
Similar to Protocol I, in the limit $\mu,\nu\rightarrow\infty$, $t_\mu,t_\nu\rightarrow0$, and implementing $\u(t,\mu,\nu)$ with the effective Hamiltonian produces
\begin{align}
    \ket{\Psi_{1}^{\text{II}}} &= \frac{1}{2}\Big(\beta\ket{M,P,0,0} + \ket{M,0,0,P} \nonumber \\
    &\qquad + \ket{0,P,M,0} - \beta\ket{0,0,M,P}\Big);\nonumber \\
    \ket{\Psi_{2}^{\text{II}}} &= \frac{1}{2}\Big(\beta\ket{M,P,0,0} + \ket{M,0,0,P} \nonumber \\
    &\qquad + i\ket{0,P,M,0} - i\beta\ket{0,0,M,P}\Big); \nonumber \\
    \ket{\Psi_{3}^{\text{II}}} &= \frac{1}{\sqrt{2}}\Big(\ket{M,P,0,0}
    + \beta e^{-i\pi/2}\ket{M,0,0,P}\Big); \nonumber \\
    \hspace{-0.4cm}\ket{\Psi_{4}^{\text{II}}} &= \frac{1}{\sqrt{2}}\Big(\ket{M,P,0,0}+ \Upsilon \ket{M,0,0,P}\Big)   \label{eq:psi4leandro}
\end{align}
where  $ \Upsilon =\beta \exp(i(P\theta - \pi/2))$. 
%
%


\subsection*{Protocol fidelities}
The analytic results provided above are obtained by employing the effective Hamiltonian in an extreme limit, with divergent applied fields acting for infinitesimally small times.
Below we give numerical simulations of the protocols to show that, for physically realistic settings where the fields are applied for finite times, high-fidelity outcomes for NOON state production persist. 

Throughout this section, we use $\ket{\Psi}$ to denote an analytic state, obtained in an idealized limit. We adopt $\ket{\Phi}$ to denote a numerically calculated state, obtained by time evolution with the EBHM Hamiltonian
(\ref{eq:extended-bh}). Two sets of parameters are chosen to illustrate the results (expressed in Hz): \\

Set 1: \{$U/\hbar = 75.876$, $J/\hbar = 24.886$, $\mu/\hbar = 20.870$\}; 

Set 2: \{$U/\hbar = 76.519$, $J/\hbar = 73.219$, $\mu/\hbar = 15.168$\}. \\

\noindent
For all numerical simulation results presented below, the initial state is chosen as $|\Psi_0\rangle=|4,11,0,0\rangle$, 
i.e. $M=4$ and $P=11$. 

The fidelities of Protocols I and II are defined as\cite{nielsen} $F_{\text{I}} = |\braket{\Psi_{3}^{\text{I}}|\Phi_{3}^{\text{I}}}|$ and $F_{\text{II}} = |\braket{\Psi_{4}^{\text{II}}|\Phi_{4}^{\text{II}}}|$, respectively. This is computed for $P\theta$ ranging from $0$ to $\pi$, achieved by varying $t_\mu$. In the case of Protocol II, we use $\nu = \mu$ for both sets of parameters. The systems considered here can, in principle, be implemented using existing hardware -- see Physical proposal. 

\begin{figure}[H]
    \centering
    \includegraphics[width=8.6cm]{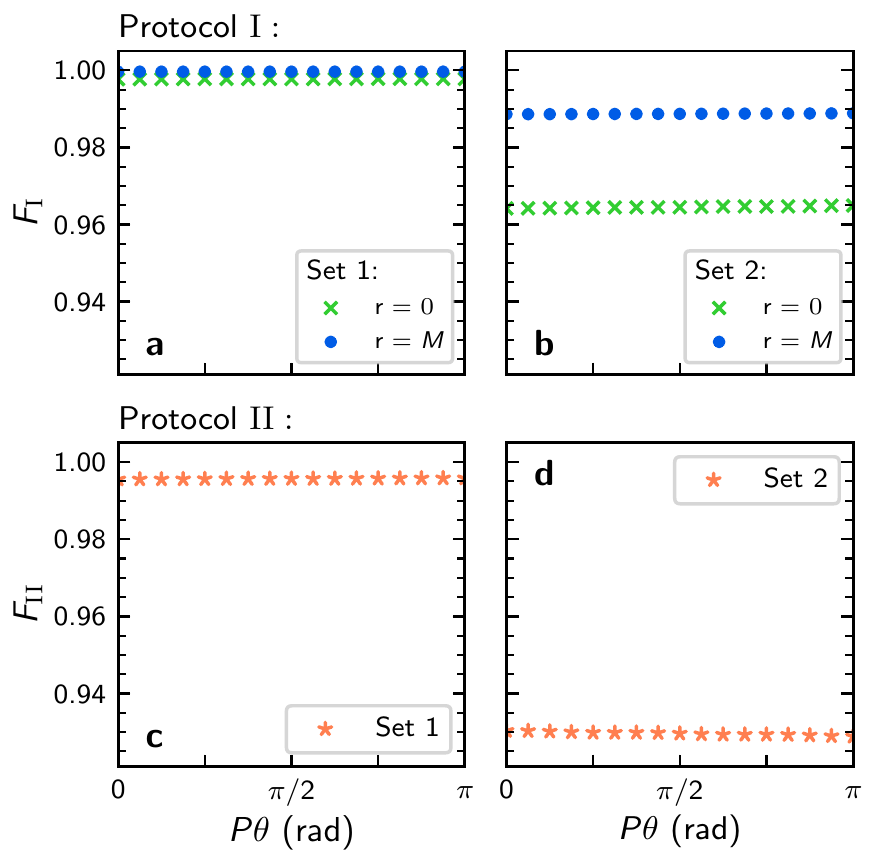}
    \caption{Fidelities for Protocols I and II. Numerical calculations of the fidelities $F_{\text{I}}$ (\textbf{a} and \textbf{b}) and $F_{\text{II}}$ (\textbf{c} and \textbf{d}). To vary $P\theta$, $\mu$ is fixed  and $t_\mu$ is varied. \textbf{a}, \textbf{c} Set 1:  $U/\hbar = 75.876$ Hz, $J/\hbar = 24.886$ Hz, $\mu/\hbar = 20.870$ Hz, 
    $t_{\text{m}} \sim 36.950$s, $t_\nu \sim 0.009$s and $t_\mu$ varies from $t=0$ to $t \simeq 0.007$s, such that $P\theta \in [0,\pi]$. \textbf{b}, \textbf{d} Set 2:   
    $U/\hbar = 76.519$ Hz, $J/\hbar = 73.219$ Hz, $\mu/\hbar = 15.168$ Hz, 
    $t_{\text{m}} \sim 4.248$s, $t_\nu \sim 0.013$s, and $t_\mu$ varies from $t=0$ to $t \simeq 0.009$s, such that $P\theta\in[0,\pi]$. 
    The required times $t_{\text{m}},\,  2t_{\text{m}}$ to produce the NOON states are comparable with typical lifetimes of optical lattice traps, which can be as large as a few minutes\cite{gibbons2008}.}
    \label{fig:fids-probs-comp}
\end{figure}

The results are presented in Fig. \ref{fig:fids-probs-comp}, where it is seen that $F_{\text{II}}$ is lower than $F_{\text{I}}$. This can be attributed to two primary causes. The first is that, while Protocol I takes $\tau_{I} \sim t_{\text{m}}$ to produce the final state, Protocol II requires double the evolution time $\tau_{II} \sim 2 t_{\text{m}}$. The longer evolution time contributes to a loss in fidelity. 
The second reason is that, the measurement occurring in the final step of Protocol I has the effect of renormalizing the quantum state after collapse, which increases the fidelity of the resulting NOON state when a measurement of $r = 0$ or $r=M$ is obtained. However, there is a finite probability that the measurement outcome is neither  $r = 0$ nor $r=M$ (see Supplementary Note 2). 

In summary, both protocols display high fidelity results greater than 0.9. For Protocol I the outcomes are probabilistic (See Supplementary Note 2 for data). By contrast, the slightly lower fidelity results of Protocol II are deterministic.

\subsection*{Readout statistics}
A means to test the reliability of the system, through a statistical analysis of local measurement outcomes, is directly built into the design. This results from the system's capacity to function as an interferometer\cite{preprint}.     
For both protocols, once the output state has been attained we can continue to let the system evolve under $\u(t_{\text{m}},0,0)$. This yields the readout states, denoted as  $\ket{\Psi_{\text{RO}}^\text{I}}$, $\ket{\Psi_{\text{RO}}^\text{II}}$ respectively for protocols I and II. In the idealized limits these are
\begin{equation*}
\ket{\Psi_{\text{RO}}^\text{I}} = \left\{
\begin{aligned}
    &\displaystyle\frac{c(\theta)}{\sqrt{2}}(\ket{M,P,0,0} + \beta\ket{M,0,0,P})\\
    &\displaystyle + \frac{is(\theta)}{\sqrt{2}}(\beta\ket{0,P,M,0} - \ket{0,0,M,P}),\text{ }r=0,\\
    &\phantom{a}\\
    &\displaystyle\frac{c(\theta)}{\sqrt{2}}(\ket{M,P,0,0} - \beta\ket{M,0,0,P})\\
    &\displaystyle - \frac{is(\theta)}{\sqrt{2}}(\beta\ket{0,P,M,0} - \ket{0,0,M,P}),\text{ }r=M,\\
\end{aligned}
\right.
\end{equation*}
\begin{flalign*}
\begin{split}
    \ket{\Psi_{\text{RO}}^{\text{II}}} &= \frac{1}{\sqrt{2}}s\left(\theta-\frac{\pi}{2P}\right)\left(\ket{M,P,0,0} + \beta\ket{M,0,0,P}\right)\\
    &\phantom{=} - \frac{i}{\sqrt{2}}c\left(\theta-\frac{\pi}{2P}\right)\left(\beta\ket{0,P,M,0} - \ket{0,0,M,P}\right),
\end{split}
\end{flalign*}
where $c(\theta) \equiv \cos\left(P\theta/2\right)$ and $s(\theta) \equiv \sin\left(P\theta/2\right)$. 
For $\ket{\Psi_{\text{RO}}^I}$, the measurement probabilities at site 3 are $\mathcal{P}(0) = \cos^2\left({P\theta}/{2}\right)$ and $\mathcal{P}(M) = \sin^2\left({P\theta}/{2}\right)$.  Combined with the probability of measuring $r = 0, M$ in step (iii), we obtain four possibilities for the total probabilities as $\mathcal{P}_{\text{I}}(0,0) = \mathcal{P}_{\text{I}}(M,0) = 0.5\cos^2\left({P\theta}/{2}\right)$ and $\mathcal{P}_{\text{I}}(0,M) = \mathcal{P}_{\text{I}}(M,M) = 0.5\sin^2\left({P\theta}/{2}\right)$. Meanwhile, for $\ket{\Psi_{\text{RO}}^{II}}$, the measurement probabilities at site 3 are $\mathcal{P}_{\text{II}}(0) = \sin^2 \left({P\theta}/{2}-{\pi}/{4}\right)$ and $\mathcal{P}_{\text{II}}(M) = \cos^2\left({P\theta}/{2}-{\pi}/{4}\right)$. As a numerical check, we consider the same sets of parameters from previous section. Then, we numerically calculate the above probabilities using the Hamiltonian (\ref{eq:extended-bh}), comparing the predicted analytic results with the numerical ones, as shown in Fig. \ref{fig:probs-readout}. 
See Supplementary Note 2 for numerical probabilities of Protocol I, and related fidelity data. For results with Set 2, see Supplementary Note 3.

\section*{Methods}

\subsection*{Resonant tunneling regime}

The Hamiltonian (\ref{eq:extended-bh}) has large energy degeneracies when $J=0$. Through numerical diagonalization of the intergable Hamiltonian for sufficiently small values of $J$, it is seen that the levels coalesce into well-defined bands, similar to that observed in an analogous integrable three-site model\cite{karin2018, arlei}.  By examination of second-order tunneling processes (see Supplementary Note 1) In this regime, an effective Hamiltonian $H_{\rm eff}$ is obtained for this regime.

\begin{figure}[H]
      \subfloat{\includegraphics[width=8.6cm]{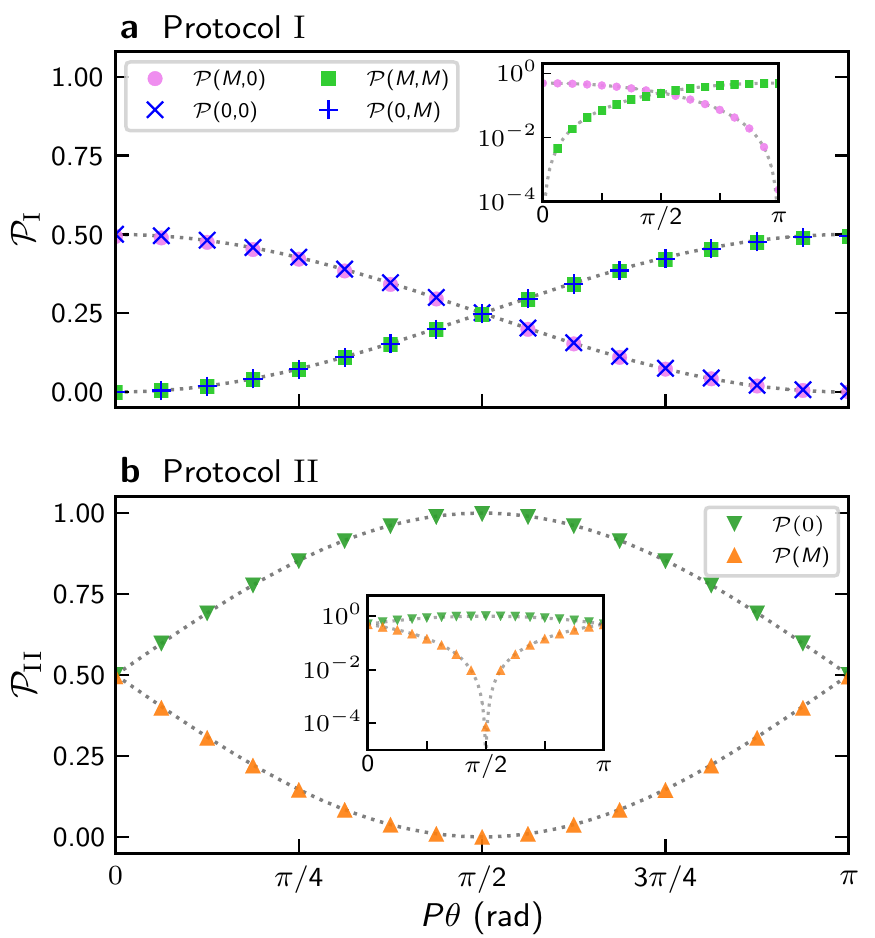}}
    \caption{Readout probabilities for Protocols I and II. Comparison between analytic and numerically-calculated probabilities for parameters of Set 1 (as in Figure \ref{fig:fids-probs-comp})
     for different values of $P\theta$. 
    {\bf a} Results for Protocol I. The pink dot and the blue ``x'' (green square and the blue ``+'') depict the probabilities of measuring $r = 0$ ($r = M$) during the readout, having measured $r = M$  or $r = 0$  in step (iii) respectively. {\bf b} Results for Protocol II. The probabilities of measuring $N_3 = 0$  ($N_3 = M$) in the readout are shown as green (orange) triangles. The dotted line depicts the analytic predictions of the probabilities with respect to $P\theta$. The insets show the accordance between predicted and calculated probabilities in semilogarithmic scale.}
    \label{fig:probs-readout}
\end{figure}

For an initial Fock state $|M-l,P-k,l,k \rangle$, with total boson number $N=M+P$, the effective Hamiltonian is a simple function of the conserved operators with the form 
\begin{equation}
    \label{hameff}
    H_{{\rm eff}} = (N+1)\Omega (Q_1 +  Q_2) -2 \Omega Q_1 Q_2,
\end{equation}
\noindent
where $\Omega=J^2/(4U((M-P)^2-1))$ and $U=(U_{12}-U_0)/4$. This result is valid for $J \ll U|M-P|$, and it is this inequality that we use to define the resonant tunneling regime. 

A very significant feature is that,  for time evolution under $H_{\rm eff}$, both $N_1+N_3=M$ and $N_2+N_4=P$ are constant. The respective $(M+1)$-dimensional subspace associated with sites 1 and 3 and $(P+1)$-dimensional subspace associated with sites 2 and 4 provide the state space for the relevant energy band (see Supplementary Note 1). Restricting to these subspaces and using the effective Hamiltonian (\ref{hameff})  yields a robust approximation for (\ref{eq:extended-bh}).

\newpage
\subsection*{Physical proposal}
We propose a physical construction, consisting of dysprosium $^{164}$Dy atoms trapped in an optical lattice, to test the theoretical results. The trapping is accomplished by employing two sets of counterpropagating laser beams with wavelength $\lambda = 0.532$ $\mu$m and waist $w_0$, with $w_0 \gg \lambda$. We consider each set of beams to cross with the other at an angle of $90^\circ$ (cyan beams in Fig. \ref{fig:high-disp}) , generating a square, two-dimensional optical lattice, in which the distance between nearest wells is $l = {\lambda}/{2} = 266$ nm. We also consider a Gaussian beam propagating towards the \textbf{z}-direction (blue beam in Fig. \ref{fig:high-disp}), with $\lambda = 0.532$ $\mu$m and waist $w_1 = 1.0$ $\mu$m, aligned to the center of a four-site square plaquette, to isolate it from the rest of the lattice. Then, to achieve a pancake-shaped trap, it is necessary to include a set of two beams with $\lambda = 0.532$ $\mu$m and waist $w_2 \sim w_0$, whose orientations are disposed at an angle of $\alpha = 60^\circ$ from each other (orange beams in Fig. \ref{fig:high-disp}), inducing a trapping aspect ratio of $\kappa^2 \equiv \omega_z/\omega_{\text{r}} = 1.464$. Together, they generate the potential $V(\rb)$:

\begin{flalign}
\begin{split}
\hspace{-0.15cm}V(\rb) &= \frac{1}{2}m \omega^2 (x^2+y^2)+\frac{1}{2}m \omega_z^2 z^2 \\
&\hspace{-0.15cm}+V_0\sin^2\left[k\left(x-\frac{l}{2}\right)\right]+V_0\sin^2\left[k\left(y-\frac{l}{2}\right)\right],
\end{split}
\label{eq:potential-V}
\end{flalign}
where $m$ is the atom's mass, $\displaystyle \omega_{\text{r}} = \sqrt{\frac{2}{m}\left(V_0 k^2 + \frac{2V_1}{w_1^2}\right)}$ and 
\begin{align*}
\omega_z &= \sqrt{\frac{2}{m}\left(\frac{\pi^2 V_2}{d_{sw}^2}+\frac{V_1}{R_1^2}\right)}, \qquad R_k=\frac{\pi w_k^2}{\lambda}
\end{align*}
are, respectively, the radial and transverse trapping frequencies. Above, $\omega = \sqrt{{4V_1}/({m w_1^2})}$ arises due to the isolation of the four-well system from the optical lattice. The values $V_1 = V_0$ and $V_2 = 9V_0$ are, respectively, the central beam's and the \textbf{x-z} crossing beam's potential depths, $V_0$ is the 2D lattice potential depth, $l$ is the distance between nearest sites, $k = 2\pi/\lambda$ is the wave number, and $d_{\text{sw}} = \lambda/\left(2 \sin\left({\alpha}/{2}\right)\right)$ is the distance between nearest wells along the \textbf{z}-axis. Since we are considering $\alpha = 60\degree$, the minimum distance between the system's horizontal layer and the next upper (or lower) layer is $d_{\text{sw}} = 2l = \lambda$, which makes irrelevant the tunneling contributions between different horizontal layers. 


To establish equivalency between $V(\rb)$ and the Hamiltonian of Eq. (\ref{eq:extended-bh}), we employ the standard second-quantization procedure. From this, we calculate the on-site interaction parameter $U_0$ as:

\begin{align}
    U_0 &=U_{\text{contact}}+U_{\text{dip}}  \nonumber    \\
    &= \frac{\kappa \eta^3}{\pi^3}\left(g - \frac{C_{\text{dd}}}{3}f(\kappa)\right),
    \label{eq:u-onsite}
\end{align}
where $\kappa$ is related to the trapping (pancake) shape aspect, $\eta \equiv m\omega_{\text{r}} / (2\hbar)$, $g\equiv 4\pi\hbar^2 a/m$, with $a$ being the s-wave scattering length (tunable via Feshbach Resonance), $C_{\text{dd}} \equiv \mu_0 \mu_1^2$ is the coupling constant, where $\mu_0$ is the vacuum magnetic permeability, $\mu_1$ is the atomic  magnetic moment, and $f(\kappa)$ is a function that describes how the dipolar interaction behaves for different geometries (encoded in $\kappa$)\cite{lahaye09}. Taking site 1 as the ``starting point', the parameter $U_{1\text{j}}$, which accounts for the dipole-dipole interaction between atoms at sites 1 and j, is expressed as:

\begin{align}
    \hspace{-.2cm}U_{1\text{j}} = &\frac{C_{\text{dd}}}{4\pi}\int_{0}^{\infty}  {\rm d}r\hspace{0.1cm} r\exp\left(-\frac{r^2}{4\eta}\right) J_0(r d_{1\text{j}}) Z(r), \label{eq:u-ddi}\\
Z(r)=    &\left(\frac{4}{3}\sqrt{\frac{\kappa^2\eta}{\pi}}  - r\exp\left( \frac{r^2}{4\kappa^2\eta}\right) \text{erfc}\left(\frac{r}{2\sqrt{\kappa^2\eta}}\right)\right),  \nonumber 
\end{align}
where $J_0$ is the Bessel function of first kind, $d_{1\text{j}} = l/\delta$, if $j = 2,4$, and $d_{1\text{j}} = l\sqrt{2}/\delta$, if $j = 3$. Here, the on-site dipolar interaction is given by $\displaystyle U_{\text{dip}} = \lim_{d_{ij}\to 0} U_{1\text{j}}\propto f(\kappa)$. 
The term 
$
 \delta = 1+2V_1/( V_0k^2 w_1^2)
$
arises when isolating the four-site region from the rest of the lattice, which causes the wells to slightly approach each other.

\subsubsection*{Integrability condition}

The physical setup above is able to simulate the EBHM. To achieve NOON-state generation, however, relies on the particular case for which the EBHM is integrable; as explained previously, this can be accomplished by making $U_0 = U_{13}$, which we call the ``integrability condition''. The approach is to first choose a value for the s-wave scattering length via Feshbach Resonance. Then, from the condition just stated, one has to adjust $\omega_{\text{r}}$ by varying the laser beams intensities\cite{bloch2005} such that, at some point, $U_0$ becomes the same as $U_{13}$. From this point every Hamiltonian parameter is evaluated only after the integrability condition is satisfied, which sets the intensity of the trapping scheme.

\begin{figure}[H]
    \centering
        \includegraphics[width=7cm]{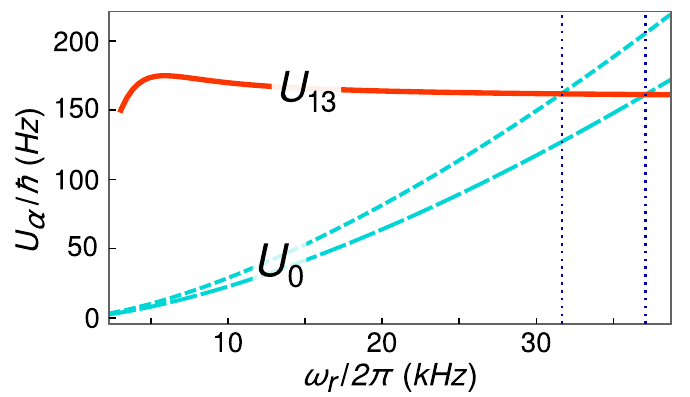}\\
    \caption{Fulfillment of integrability condition. The s-wave scattering length value is set, following by a variation of the radial trapping frequency $\omega_{\text{r}}$ up to the point at which $U_0 = U_{13}$, corresponding to the frequency required for the system to be integrable. The long dashed, short dashed and solid lines depict $U_0$ for $a = -21$ $a_0$ and $a = -20.85$ $a_0$ and $U_{13}$, respectively, for different values of $\omega_{\text{r}}$. By setting $a = -21$ $a_0$ ($-20.85$ $a_0$), we find $\omega_{\text{r}} \approx 2\pi \times 37.078$ kHz ($31.610$ kHz ) as the frequency for integrability, which results in $U_0/\hbar = U_{13}/\hbar \approx 161.282$ Hz ($161.797$ Hz). The points where $U_{0} = U_{13}$ and the corresponding frequencies $\omega_{\text{r}}$ are highlighted by the dotted lines. The system is robust for small deviations from the integrable point (see Supplementary Note 4 for more details).}
    \label{fig:int-condition}
\end{figure}

\begin{figure}[H]
    \centering
    \includegraphics[width=15cm]{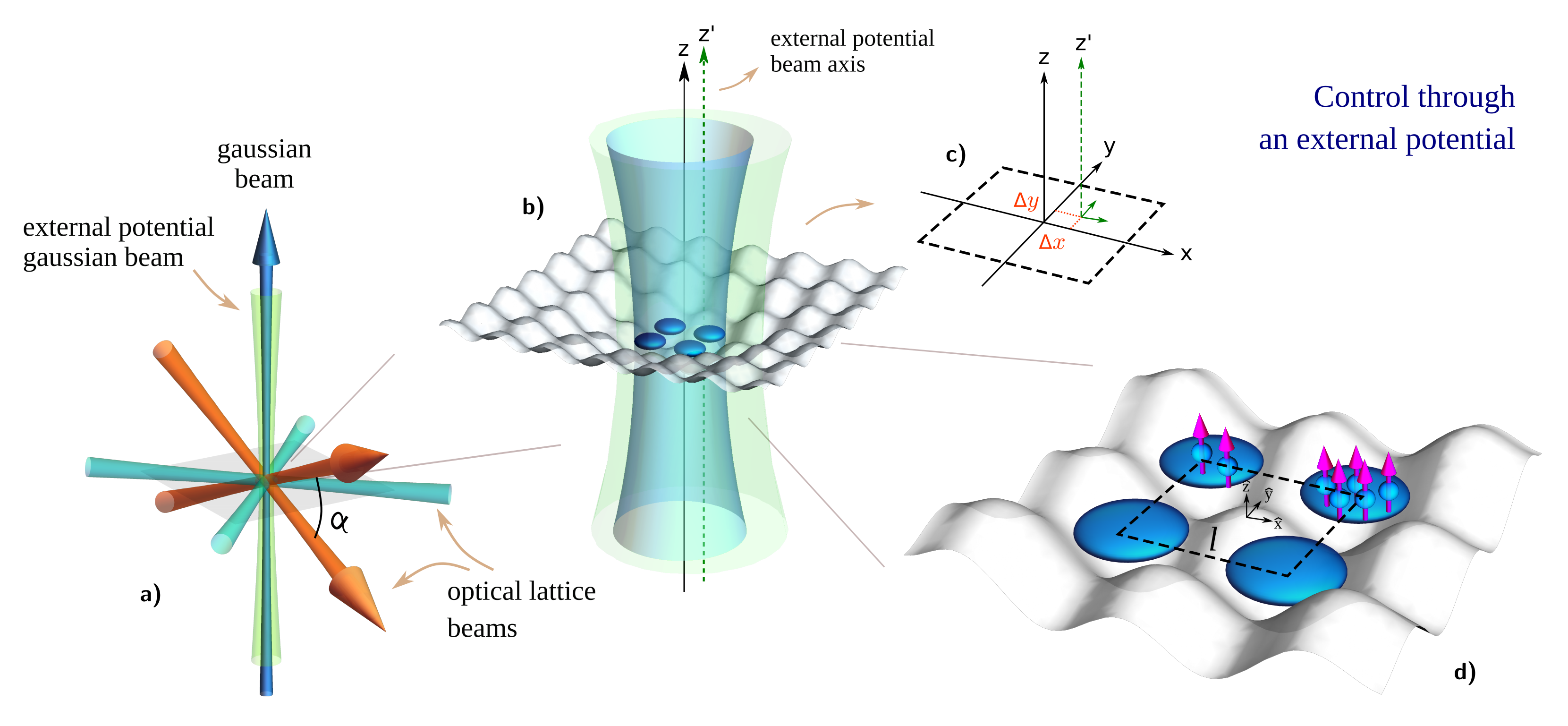}
    \caption{Representation of the trapping scheme. \textbf{a} Trapping scheme of the four-well model. In cyan, the two sets of counterpropagating beams are represented, with each set crossing at 90$\degree$ with the other, providing the two-dimensional lattice trap. In orange, the two beams crossing at an angle of $\alpha = 60\degree$ are depicted, whose propagation occur in opposite orientations as seen with respect to 
    \textbf{z}-axis, resulting in the pancake-shaped potential. In blue, the single laser beam is illustrated, whose waist value is at the typical size of the four-site system, isolating it from the rest of the 2D lattice. The external beam, used for breaking the system's integrability, is depicted in green. \textbf{b} Zoom into the region of the lattice which contains the four-well system. The blue beam represents the single laser beam isolating the region of interest from the rest of the lattice. The green beam depicts the external beam, used to controllably break the system's integrability, and the four pancakes illustrate the four wells of the system. \textbf{c} The dashed square in the \textbf{x-y} plane illustrates the square plaquette formed by the four-well system. The displacement of the central position of the green beam with respect to the center of the four-well system is represented by $\Delta x$ and $\Delta y$, which implement the breaking of the system's integrability. \textbf{d} The light grey background represents the trapping potential in the vicinities of the four-well system. The four pancake-shaped wells, at a distance of $l$ between nearest neighbors, are depicted in blue, the cyan spheres illustrate the trapped atoms and the purple arrows represent the aligned dipoles, which induce the dipole-dipole interaction.}
    \label{fig:high-disp}
\end{figure}

By considering $a = -21$ ($-20.85$) $a_0$, the system becomes integrable at $\omega_{\text{r}} \approx 2\pi\times 37.078$ ($2\pi \times 31.610$) kHz, as is depicted in Fig. \ref{fig:int-condition}. This frequency implies on a 2D-lattice depth of $V_0 \approx 18.495 E_{\text{R}}$ ($13.443 E_{\text{R}}$), where $E_{\text{R}}/\hbar = \hbar(k\pi)^2/(2ml^2) = 26.894$ kHz is the recoil energy, which characterizes a deep lattice. This allows for a higher stabilization of the system with a negative value for the s-wave scattering length\cite{muller2011}. Then, by using this trapping frequency to calculate the Hamiltonian parameters, one finds $U/\hbar \approx 75.876$ ($76.519$) Hz and $J/\hbar \approx 24.886$ ($73.219$) Hz. It is also important to highlight that the tunneling parameter $J_{13}$ between diagonal sites (1-3 and 2-4), which is not included in the Hamiltonian (\ref{eq:extended-bh}), is very small if compared to $J$. From this, one infers that the tunneling between different horizontal layers of the optical lattice is even smaller, since the distance between these layers is bigger than the distance between diagonal sites by a factor of $\sqrt{2}$.

\subsubsection*{Breaking of integrability}

To produce a controllable breaking of integrability, it is sufficient to consider a second \textbf{z}-oriented Gaussian beam (green beam in Fig. \ref{fig:high-disp}), weaker than the one used for the region isolation, with waist $w_{\text{b}} \sim 5$ $\mu$m and wavelength $\lambda = 0.532$ $\mu$m. This beam is displaced by $\Delta x$ and $\Delta y$ (with $|\Delta x|$=$|\Delta y|$) from the center of the four-well system. When the laser is turned on, it implements the terms     

\begin{flalign}
\begin{split}
    &\nu = \frac{2V_{\text{b}} l}{w_{\text{b}}^2 \delta}(\Delta x + \Delta y),\qquad 
    \mu = \frac{2V_{\text{b}} l}{w_{\text{b}}^2 \delta}(\Delta x - \Delta y),\\
\end{split}
\label{eq:mu-nu-breaks}
\end{flalign}
where $V_{\text{b}} = 5\times 10^{-3} V_0$ is the potential depth generated by the second beam. For $|\Delta x| = |\Delta y| = 0.2$ $\mu$m and the previously obtained radial trapping frequency, the parameters $\mu$ and $\nu$ can (non-simultaneously) assume the value of $20.870$ $(15.168)$ Hz. Therefore, considering $M = 4$ and $P = 11$, one should vary $t_\mu$ from $0$ to $\sim 0.007$ $(0.009)$ s to encode $P\theta$ from $0$ to $\pi$. Also, from the condition $2 \nu t_\nu/\hbar = \pi/(2M)$, $t_\nu \sim 0.009$ $(0.013)$s.

An alternative physical setup is one designed to generate many copies of disconnected four-site plaquettes. This can be realized by overlapping two square optical lattices, 
each one with lattice spacing determined through different wavelengths ($\lambda$ and $2 \lambda$) \cite{bing}.
 By changing the relative phase, the breaking of integrability can be simultaneously controlled in all copies of the four-site plaquettes. 
 
\section*{Discussion}
We have offered new techniques to address the highly challenging problem of designing a framework to facilitates NOON state creation. 
Our approach employs dipolar atoms confined to four sites of an optical lattice. The setup allows for the interactions to be tuned, and to fix the couplings in such a way that the system is integrable. At these couplings, and for controlled perturbative breaking of the integrability, the theoretical properties of the system become very transparent. 

The insights gained from integrability allowed us to develop two protocols. Protocol I employs a local measurement procedure to produce NOON states with slightly higher fidelities, over a shorter time, than Protocol II. However Protocol I is probabilistic, requiring post-selection on the measurement outcome. This is in contrast  to the deterministic   approach of Protocol II.  For both protocols,  phase-encoding is performed by breaking the system's integrability, in a controllable fashion, at specific moments during the time evolution. And in  both protocols the output states were shown to have high-fidelity in numerical simulations. We also identifed a readout scheme,  by converting encoded phases into a population imbalance, that allows verification of NOON state production through measurement statistics.   

The approaches we have described, that are based on the formation of an uber-NOON state en route to the final state, have two significant advantages. One is that the evolution time does not scale with the total number of particles. Instead, it is only dependent on the difference in particle number of subsystem A and B in the Fock-state input. The other advantage is that all measurements are made in the local Fock-state basis.  

We conducted an analysis of the feasibility of a physical proposal. It was demonstrated that the long-range interaction between dipolar atoms allows for an integrable coupling to be achieved, depending on the interplay between contact and dipolar interactions. Through the second-quantization procedure the values for the Hamiltonian parameters were provided, derived by numerical calculations. These are seen to be realistic both in the context of optical-lattice setups and in comparison to literature. We also outlined a procedure to improve the system's robustness with respect to error perturbation (see Supplementary Note 4 for a broader description). 

Besides demonstrating the feasibility of NOON state generation, the physical setup we provide can also be employed in the study of thermalization processes and other many-body features of the EBHM. By establishing a link between integrability and quantum technologies, this work promotes advances in the field of neutral-atom quantum information processing.  

\end{multicols}

\begin{multicols}{2}

\section*{Data availability}
All relevant data are available on reasonable request from the authors.


\section*{Acknowledgements}

D.S.G. and K.W.W. were supported by CNPq (Conselho Nacional de Desenvolvimento Cient\'{\i}fico e Tecnol\'ogico), Brazil. A.F. acknowledges support from CNPq - Edital Universal 430827/2016-4. A.F. and J.L. received funding from the Australian Research Council through Discovery Project DP200101339. J.L. acknowledges the traditional owners of the land on which The University of Queensland operates, the Turrbal and Jagera people. We thank Ricardo R. B. Correia and Bing Yang for helpful discussions. 

\section*{Author contributions} 

All authors contributed to the conceptualization of the project, and actively engaged in the writing of the manuscript. D.S.G, K.W.W. and L.H.Y. implemented the theoretical analyses of the model, detailed the physical proposal, and processed the numerical computations. J.L. and A.F. designed the research framework, and directed the program of activities.

\newpage
\appendix

\setcounter{page}{1}
\setcounter{equation}{0}
\renewcommand{\theequation}{S.\arabic{equation}}
\setcounter{figure}{0}
\renewcommand\figurename{\bf Supplementary Figure}
\flushbottom
\subsection*{Supplementary Note 1: 
Energy bands and effective Hamiltonian}
Here we give an overview of the origin for the effective Hamiltonian. Recall that the integrability condition is $U_{13}=U_{24}=U_0$ and $U_{12}=U_{23}=U_{34}=U_{14}$. When $J=0$, the Fock state $\ket{M-l,P-k,l,k}$ is eigenstate of the Hamiltonian (\ref{eq:extended-bh}) with energy 
\begin{align}
E&=C    -U(M-P)^2  \label{nrg}
\end{align}
where $C=(U_0+U_{12})N^2/{4}-{U_0}/{2}$. The result is  
independent of $l$ and $k$, indicating  degeneracies. For small values of $J$, the degeneracies are broken and lead to energy levels in well-defined bands, each with $2(M+1)(P+1)$ energy levels, except for $N$ even, where the band with the highest energy, $M=P$, will have $(M+1)(P+1)$ levels. The level energy structure of the case we are analyzing, with $N = 15$, is shown in Supplementary Figure 1. In it, we highlight in cyan the band with $M = 4$ and $P = 11$ (and vice versa), while the vertical lines marks the two sets of parameters pointed in the main text (repeated here, expressed in Hz):\\

Set 1: \{$U/\hbar = 75.876$, $J/\hbar = 24.886$, $\mu/\hbar = 20.870$\};

Set 2: \{$U/\hbar = 76.519$, $J/\hbar = 73.219$, $\mu/\hbar = 15.168$\}.

\begin{figure}[H]
\begin{center}  
    \includegraphics[width=8cm]{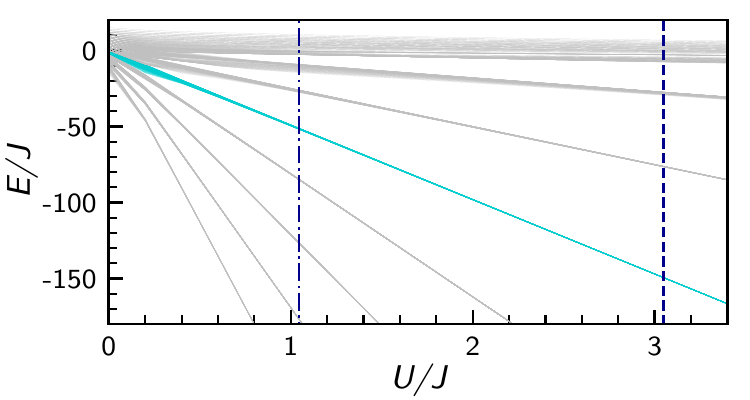}
    \caption{Energy band formation. Dimensionless energy eigenvalues $E/J$ as a function of dimensionless coupling $U/J$, where $U=(U_{12}-U_0)/4$ and considering $C=0$ in (\ref{nrg}).
    The dashed vertical line marks $U/J \sim 3 $ (concerning parameter Set 1) and the dot-dashed line marks ($U/J \sim 1$) (concerning parameter Set 2), while cyan depicts the band containing the expectation energy of the initial state $|\Psi_0\rangle=|4,11,0,0\rangle$. The formation of the bands is due to the quadratic dependence of $(M-P)$ in the energy (\ref{nrg}).}
\end{center}
\label{bands}
\end{figure}
An effective Hamiltonian for each band is obtained by consideration of second-order processes. Associated to labels $M$ and $P$, such that $N=M+P$, we obtain
\begin{align*}
\hspace{-0.0cm}H_{\rm eff}&=\frac{J^2}{16U(M-P+1)}
\left(a_1 a_3^\dagger+a_3 a_1^\dagger  \right)\left(a_2^\dagger a_2+a_4^\dagger a_4  \right)\\
& +\frac{J^2}{16U(M-P+1)}
\left(a_1 a_1^\dagger+a_3 a_3^\dagger   \right)\left(a_2^\dagger a_4 +a_4^\dagger a_2  \right)\\
& -\frac{J^2}{16U(M-P-1)}
\left(a_2 a_2^\dagger+a_4 a_4^\dagger  \right)
\left(a_1^\dagger a_3 +a_3^\dagger a_1  \right)
\\
& -\frac{J^2}{16U(M-P-1)}
\left(a_2 a_4^\dagger+a_4 a_2^\dagger  \right)
\left(a_1^\dagger a_1+a_3^\dagger a_3   \right)
\\
& +\frac{J^2}{16U}\left(\frac{1}{M-P+1}-\frac{1}{M-P-1}\right) \\
&\times \left(a^\dagger_1a_2a_3a^\dagger_4+  a^\dagger_1a^\dagger_2a_3a_4 + a_1a^\dagger_2a^\dagger_3a_4 + a_1a_2a^\dagger_3a^\dagger_4\right) .
\end{align*}		
For a given initial Fock state, the resonant regime is achieved when the expectation energy lies in a region characterized by an energy band. There, the values of the integrability-breaking parameters $\mu$, $\nu$ may be as large as the band-separation allows, which is depicted in Supplementary Figure 2.

\begin{figure}[H]
    \centering
        \subfloat{\includegraphics[width=8cm]{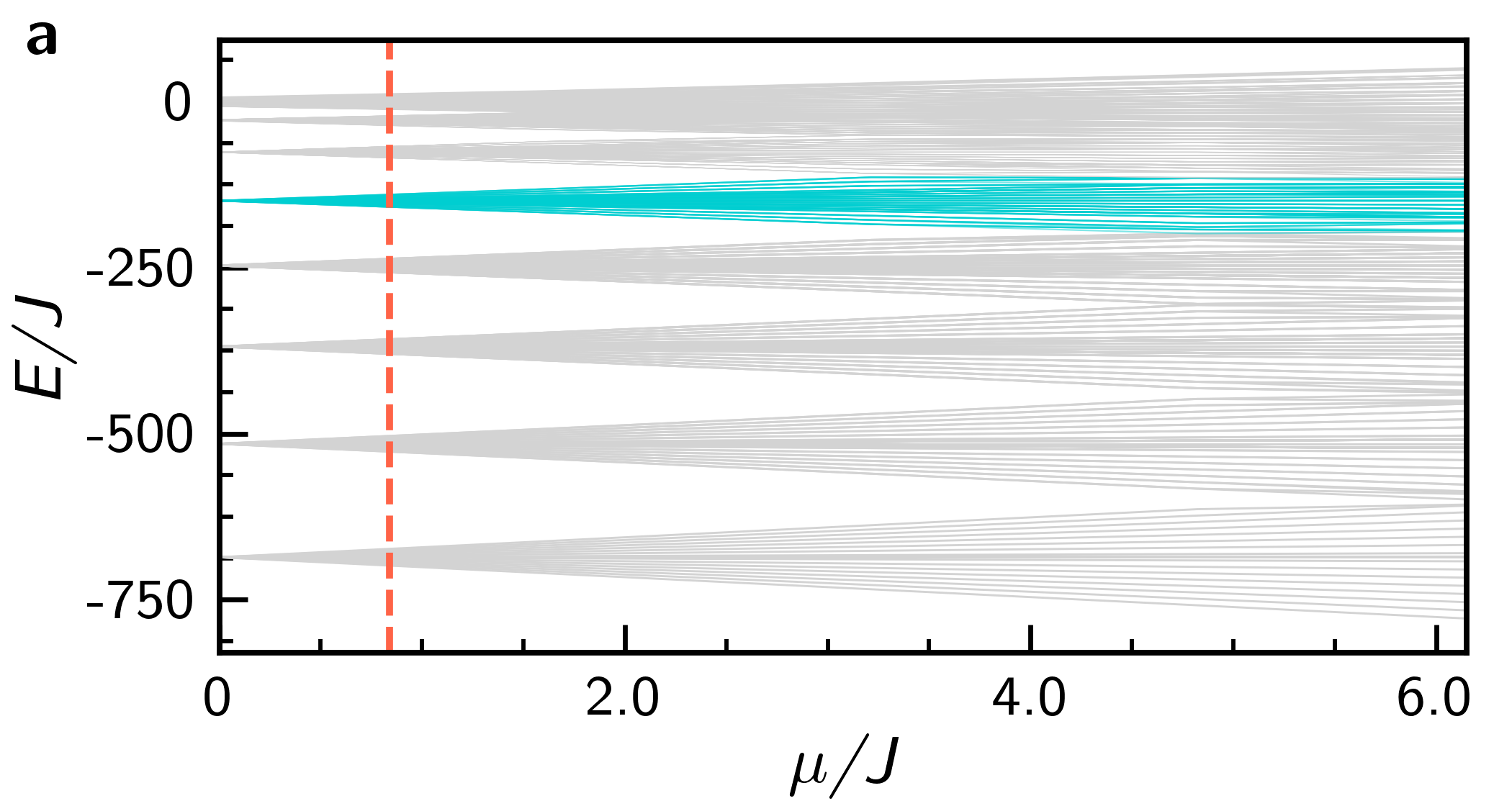}}
                \vspace{-0.4cm}
                \subfloat{\includegraphics[width=8cm]{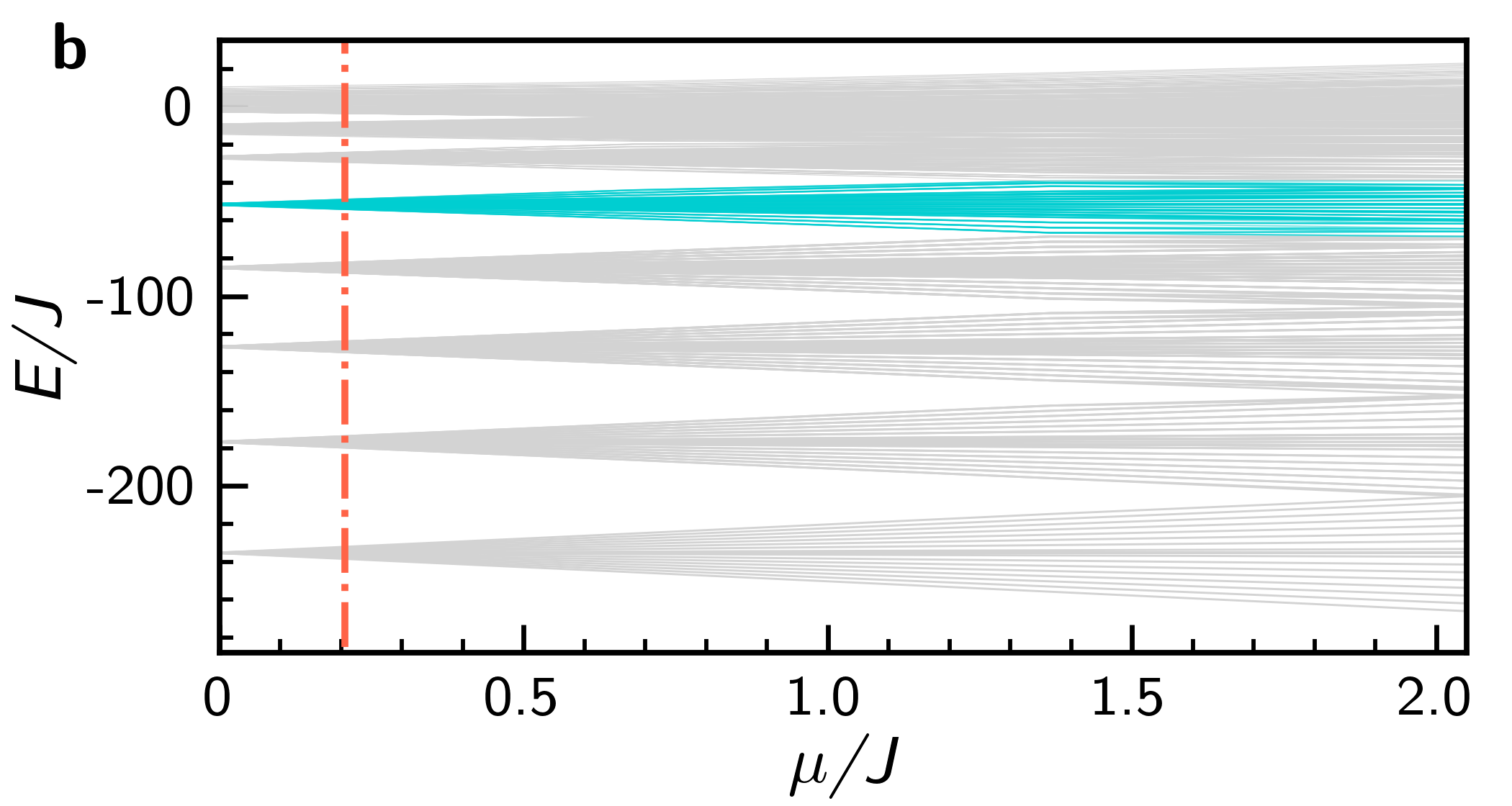}}
    \caption{Energy bands for broken integrability. Four-well model energy distribution for the two sets of parameters $U$ and $J$, as in the main text. {\bf a} 
    Set 1: $U/J \sim 3$ and {\bf b}  Set 2: $U/J \sim 1$.
    The vertical lines indicate the respective integrability-breaking parameters
    $\mu/\hbar = 20.870$ Hz ({\bf a}) and $\mu/\hbar = 15.168$ Hz ({\bf b}). The cyan lines represent the energy band associated to the initial state $\ket{\Psi_0} = \ket{4,11,0,0}$.}
    \label{fig:bands-break}
\end{figure}

\subsection*{Supplementary Note 2: 
Probabilities and fidelities}
Supplementary Table 1 shows the measurement probabilities of Protocol I, as well as the fidelity of the resulting state with the respective NOON state, for $M = 4$, $P = 11$ and the two aforementioned sets of parameters. The resulting NOON state from Protocol I can be either symmetric ($r = 0$) or antisymmetric  ($r = M$). For intermediate values for the outcome of measuring $N_3$, we calculate the fidelity of the resulting state with the symmetric NOON state ($r = 0,1$) or the antisymmetric state ($r = 2, 3, 4$), respectively.

\end{multicols}


\setcounter{equation}{0}
 \renewcommand{\tablename}{\bf Supplementary Table}

 \begin{table}[h!]
 \label{tab:probs-prot1-1}
 \noindent{\bf\normalsize Protocol I:}
 \begin{center}
\def\arraystretch{1.3}   
\resizebox{16cm}{!}{
    \begin{tabular}{c|cc|cc|cc|cc|cc|cc}
     \multicolumn{1}{c}{} & \multicolumn{12}{l}{\bf\large Set 1}\\
  \cline{2-13}
 \multicolumn{1}{c}{} & \multicolumn{12}{c }{Phase ($P\theta$)}\\
      \hline
 \multicolumn{1}{c}{Measurement}  & \multicolumn{2}{c}{$0$} & \multicolumn{2}{c}{$\pi/6$} & \multicolumn{2}{c}{$\pi/4$} & \multicolumn{2}{c}{$\pi/3$} & \multicolumn{2}{c}{$\pi/2$} & \multicolumn{2}{c}{$\pi$} \\
    \hline
  \bf{$r$} & $\mathcal{P}(r)$  & $F_\text{I}$
            & $\mathcal{P}(r)$  & $F_\text{I}$
            & $\mathcal{P}(r)$  & $F_\text{I}$
            & $\mathcal{P}(r)$  & $F_\text{I}$
            & $\mathcal{P}(r)$  & $F_\text{I}$
            & $\mathcal{P}(r)$  & $F_\text{I}$\\
    \hline
   0 & 0.5009 & 0.9977 
        & 0.5009 & 0.9977 
        & 0.5009 & 0.9978 
        &  0.5009 & 0.9978 
        &  0.5009 & 0.9977 
        & 0.5009 & 0.9978\\
    
    1 & 0.0006 & 0.0488 
        & 0.0006 & 0.0489  
        & 0.0006  &  0.0494  
        & 0.0006  &  0.0499 
        & 0.0006  &  0.0501 
        &  0.0006  &  0.0512 \\
  
    2  & 0.0003  &  0.0164 
        & 0.0003  &  0.0160 
        & 0.0003  &  0.0155 
        & 0.0003  &  0.0162 
        & 0.0003  &  0.0169 
        & 0.0003  &  0.0155\\
    
    3  & 0.0013  &  0.0447 
        & 0.0013  &  0.0450  
        & 0.0013  &  0.0452  
        & 0.0013  &  0.0453  
        & 0.0013  &  0.0454 
        & 0.0013  &  0.0463\\
    
    \textit{M}=4  & 0.4956  &  0.9996 
        & 0.4957  &  0.9996 
        & 0.4956  &  0.9996  
        & 0.4956  &  0.9996 
        & 0.4957  &  0.9996 
        & 0.4957  &  0.9996\\
   \hline
    \end{tabular}}
    \end{center}
 \begin{center}
  \vspace{0.1cm}
\def\arraystretch{1.3}   
\resizebox{16cm}{!}{
    \begin{tabular}{c|cc|cc|cc|cc|cc|cc}
\multicolumn{1}{c}{} & \multicolumn{12}{l}{\bf\large Set 2}\\
  \cline{2-13}
 \multicolumn{1}{c}{} & \multicolumn{12}{c }{Phase ($P\theta$)}\\
      \hline
 \multicolumn{1}{c|}{ Measurement}  & \multicolumn{2}{c|}{$0$} & \multicolumn{2}{c|}{$\pi/6$} & \multicolumn{2}{c|}{$\pi/4$} & \multicolumn{2}{c|}{$\pi/3$} & \multicolumn{2}{c|}{$\pi/2$} & \multicolumn{2}{c}{$\pi$} \\
    \hline
  $r$ & $\mathcal{P}(r)$  & $F_\text{I}$
            & $\mathcal{P}(r)$  & $F_\text{I}$
            & $\mathcal{P}(r)$  & $F_\text{I}$
            & $\mathcal{P}(r)$  & $F_\text{I}$
            & $\mathcal{P}(r)$  & $F_\text{I}$
            & $\mathcal{P}(r)$  & $F_\text{I}$\\
    \hline
   0 & 0.4922 & 0.9642 
        & 0.4922 & 0.9643 
        & 0.4922 & 0.9644 
        &  0.4922 & 0.9644 
        &  0.4922 & 0.9645 
        & 0.4923 & 0.9649\\
    
    1 & 0.0097 & 0.1219 
        & 0.0097 & 0.1221  
        & 0.0097  &  0.1219  
        & 0.0097  &  0.1214 
        & 0.0097  &  0.1221 
        &  0.0096  &  0.1214 \\
  
    2  & 0.0053  &  0.0400 
        & 0.0053  &  0.0384 
        & 0.0053  &  0.0378 
        & 0.0053  &  0.0375 
        & 0.0053  &  0.0363 
        & 0.0053  &  0.0320\\
    
    3  & 0.0139  &  0.1336 
        & 0.0139  &  0.1338  
        & 0.0139  &  0.1333  
        & 0.0139  &  0.1332  
        & 0.0139  &  0.1332 
        & 0.0139  &  0.1325\\
    
    \textit{M}=4  & 0.4629  &  0.9886 
        & 0.4631  &  0.9886 
        & 0.4632  &  0.9887  
        & 0.4633  &  0.9887 
        & 0.4635  &  0.9887 
        & 0.4640  &  0.9888\\
   \hline
    \end{tabular}}
    \vspace{0.2cm}
\caption{Measurement probabilities and NOON state fidelities. Probability of measuring $r$ particles at site 3 of Protocol I, and fidelity of the resulting state with the symmetric NOON state ($r = 0,1$) or the antisymmetric NOON state ($r = 2,3,4$). In this calculation, we employed the parameters Set 1 and Set 2 and considered $M = 4$ and $P = 11$.}
\label{tab:probs-prot1-2}
\end{center}
\end{table}

\begin{multicols}{2}
\setcounter{figure}{2}
\renewcommand\figurename{\bf Supplementary Figure}
\setcounter{equation}{1}
\renewcommand{\theequation}{S.\arabic{equation}}

\subsection*{Supplementary Note 3: Readout statistics}
For less ideal choices of parameters, it is possible to perform a fitting on the readout probabilities amplitudes, such that
\begin{equation}\label{fit}
\begin{aligned}
&\text{Protocol I}
\begin{cases}
\displaystyle \mathcal{P}(0,0) = \mathcal{P}(M,0) = \frac{c_{00}}{2}\cos^2\left(\frac{P\theta}{2}\right)\\
\displaystyle \mathcal{P}(0,M) = \mathcal{P}(M,M) = \frac{c_{MM}}{2}\sin^2\left(\frac{P\theta}{2}\right)
\end{cases}
\\\\
&\text{Protocol II}
\begin{cases}
\displaystyle \mathcal{P}(0) = c_0\sin^2\left(\frac{P\theta}{2}-\frac{\pi}{4}\right)\\ 
\displaystyle \mathcal{P}(M) = c_M\cos^2\left(\frac{P\theta}{2}-\frac{\pi}{4}\right)
\end{cases}
\end{aligned}
\nonumber
\end{equation}
where $c_{00}$, $c_{MM}$, $c_{0}$ and $c_{M}$ are constants that are obtained by fitting the numerically-evaluated data with the analytic models. By choosing the parameters of Set 2, we obtain the following constants from a least-squares fitting: $c_{00} = 0.938$, $c_{MM} = 0.893$, $c_0 = 0.954$ and $c_M = 0.909$. The results are shown in Supplementary Figure 3.

\begin{figure}[H]
    \subfloat{\includegraphics[width=8.6cm]{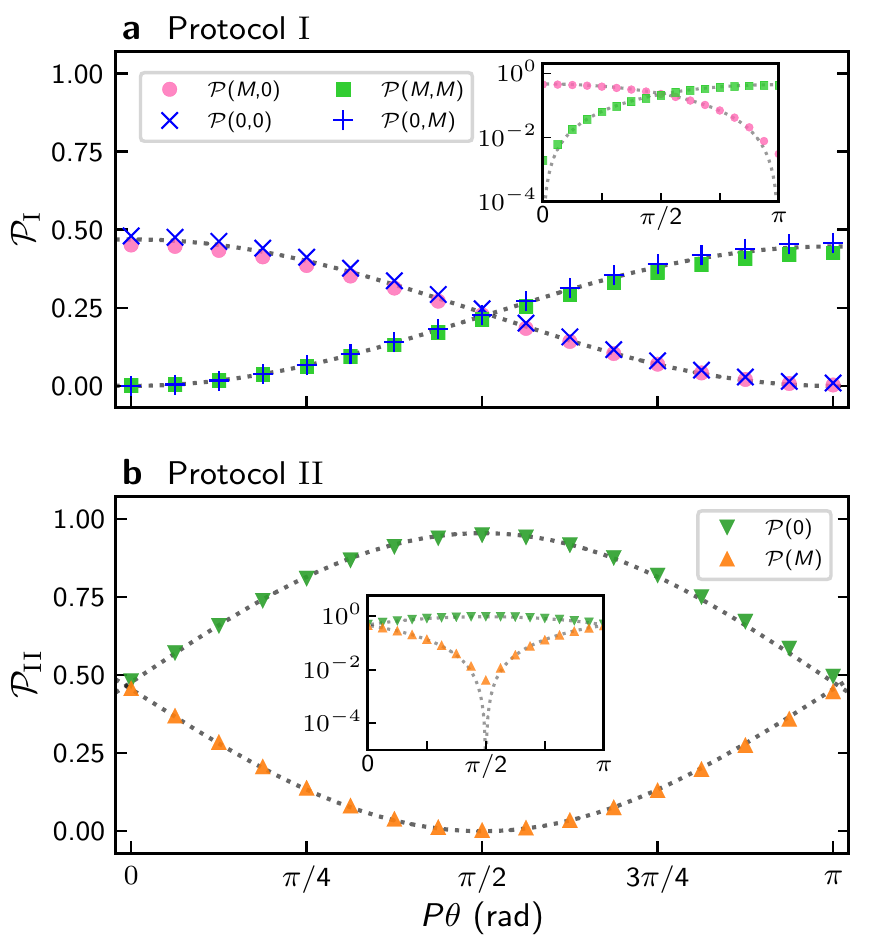}}
\caption{Readout probabilities. Comparison between analytic and numerically-calculated probabilities relating to the parameters of Set 2 
with $\mu/\hbar = \nu/\hbar = 15.168$ Hz, for different values of $P\theta$. {\bf a)} Probability distributions for measuring $N_3 = 0$ ($N_3 = M$) after time evolution subsequent to Protocol I. {\bf b)} Probability distributions of measuring $N_3 = 0$ ($N_3 = M$) after time evolution subsequent to Protocol II. In both cases, the dotted lines refer to the analytic probabilities adjusted to the numerical points according to the Eqs. \eqref{fit}. The coefficients used were $c_{00} = 0.938$, $c_{MM} = 0.893$, $c_0 = 0.954$ and $c = 0.909$. The insets show the accordance between predicted and calculated probabilities in semilogarithmic scale.}
    \label{fig:probs-jon-appendix}
\end{figure}


\setcounter{figure}{3}
\renewcommand\figurename{\bf Supplementary Figure}

\subsection*{Supplementary Note 4: Robustness}

Here we analyze the system's robustness in the presence of a perturbation parameter, and outline a method to enhance performance. Supposing that the integrability condition is subject to an error, denoted by $\xi$:
\begin{equation}
\xi =    U_{0} - U_{13}.
    \label{eq:appendix-int-condition}
\end{equation}
We find that the fidelities for the parameters Set 1 are above 0.9 for an error parameter $\xi/J$ up to $\sim 0.01\%$, while the parameter Set 2 is able to produce NOON state with fidelities above $0.9$ up to $\xi/J \sim 0.03\%$.

 To enhance the fidelity, we propose a procedure that consists of both positive ($+\xi$) and negative ($-\xi$) deviations in Eq. (\ref{eq:appendix-int-condition}). This can be done, for instance, by considering a sequence of pulses\cite{zhou2018}. This is appropriate when considering an error parameter in the physical setup: after fixing the desired (approximate) s-wave scattering length, the trapping frequency adjustment may not have the required precision, allowing for a minimum-error of $\pm\xi$. 
 
Considering  perturbations  $H^{\pm}(\mu,\nu)$ of the Hamiltonian, with the form
\begin{flalign}
\begin{split}
    H^{\pm}(\mu,\nu) &= H(U^\pm, J^\pm) + \mu^{\pm}(N_2 - N_4)\\
    &\hspace{-0.3cm} + \nu^{\pm}(N_1 - N_3) \pm \xi(N_1 N_3 + N_2 N_4),
\end{split}
\nonumber
\end{flalign}
set $\Bar{U}$, $\Bar{J}$ and $\Bar{\mu}$ as the mean values for the two cases $+\xi$ and $-\xi$.  We then calculate the times $t_\text{m}$ and $t_\mu$ from these mean values. Next, running a simulation that alternates $N_{\delta t}$ times between the extreme coupling values during the integrable time evolution over $t_\text{m}- t_\mu$ leads to an increase in the system's tolerance to the error, as depicted in Supplementary Figure 4.

\begin{figure}[H]
    \centering
        \subfloat{\includegraphics[width=8.6cm]{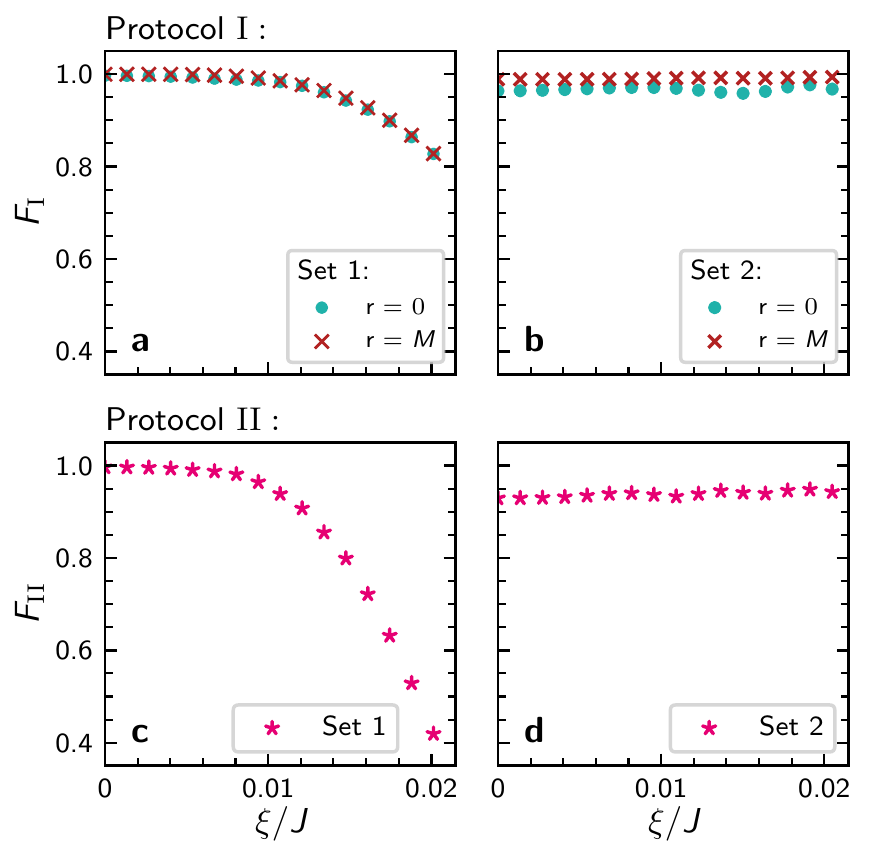}}
\caption{Robustness. NOON states fidelities with respect to a perturbation parameter $\xi$, for Protocols I (panels \textbf{a} and \textbf{b}) and II (panels \textbf{c} and \textbf{d}). Panels \textbf{a} and \textbf{c}: Set 1.
Panels \textbf{b} and \textbf{d}: 
Set 2. 
We considered $\nu = \mu$ in Protocol II evaluations. For every $\xi$, we evaluate the fidelities, with $P\theta=\pi/2$, for the Hamiltonian parameters obtained by solving for $\omega_{\text{r}}$ that corresponds to Eq. (\ref{eq:appendix-int-condition}) (cf. Figure \ref{fig:int-condition} of the main text). With $N_{\delta t} = 100$ oscillations between $+\xi$ and $-\xi$, NOON states are produced with fidelities higher than 0.9 for $\xi/J$ up to $1.2\%$ on the left, and more than $2.0\%$ on the right.}
    \label{fig:fids-robust-action}.
\end{figure}

\renewcommand{\refname}{\bf Supplementary Reference}

\end{multicols}
\end{document}